\documentclass{emulateapj} \usepackage{apjfonts}
\newcommand\cs{c_s}

\newcommand\vA{v_{\rm A}}

\newcommand\kms{{\;\rm km\; s^{-1}}}
\newcommand\torb{t_{\rm orb} }

\newcommand\kpc{{\;\rm kpc}}
\newcommand\simgt{\lower.5ex\hbox{$\; \buildrel > \over \sim \;$}}
\newcommand\simlt{\lower.5ex\hbox{$\; \buildrel < \over \sim \;$}}

\shorttitle{ISM TURBULENCE DRIVING BY SPIRAL SHOCKS}
\shortauthors{KIM, KIM, \& OSTRIKER}
\begin{document}

\title{Interstellar Turbulence Driving by Galactic Spiral Shocks}

\author{Chang-Goo Kim\altaffilmark{1}, Woong-Tae Kim\altaffilmark{1},
and Eve C.\ Ostriker\altaffilmark{2}}
\affil{$^1$Department of Physics \& Astronomy, FPRD,
Seoul National University, Seoul 151-742, Republic of Korea}
\affil{$^2$Department of Astronomy, University of Maryland, 
College Park, MD 20742, USA}
\email{kimcg@astro.snu.ac.kr, wkim@astro.snu.ac.kr, ostriker@astro.umd.edu}

\begin{abstract}
  Spiral shocks are potentially a major source of turbulence in the
  interstellar medium.  To address this problem quantitatively, we use
  numerical simulations to investigate gas flow across spiral arms in
  vertically stratified, self-gravitating, magnetized models of
  galactic disks.  Our models are isothermal, quasi-axisymmetric, and
  local in the quasi-radial direction while global in the vertical
  direction.  We find that a stellar spiral potential perturbation
  promptly induces a spiral shock in the gas flow. For vertically
  stratified gas disks, the shock front in the radial-vertical plane
  is in general curved, and never achieves a steady state. This
  behavior is in sharp contrast to spiral shocks in two-dimensional
  (thin) disks, which are generally stationary.  The non-steady
  motions in our models include large-amplitude quasi-radial flapping
  of the shock front.  This flapping feeds random gas motions on the
  scale of the vertical disk thickness, which then cascades to smaller
  scales. The induced gas velocity dispersion in quasi-steady state
  exceeds the sonic value for a range of shock strengths, suggesting
  that spiral shocks are indeed an important generator of turbulence
  in disk galaxies.
\end{abstract}

\section{INTRODUCTION}

Turbulence in the interstellar medium (ISM) is observed to be
pervasive and highly supersonic.  Shocks created by random gas motions
produce a rich variety of structures in the diffuse, atomic ISM as
well as in gravitationally-bound giant molecular clouds (GMCs).
Recent work has shown that turbulence is crucial to control of star
formation within GMCs, and it is also believed to affect GMC formation
processes. Turbulence thus regulates star formation on both
local and global scales (e.g., \citealt{bal06}).  In the absence of
driving, shock dissipation and nonlinear cascades cause turbulence to
decay on a time scale comparable to flow crossing times even for a
medium with equipartition-strength magnetic fields (e.g.,
\citealt{sto98,mac99,pad99}), amounting to a few tens of Myr for
the diffuse ISM.  This implies that the ISM must be continuously
stirred by one or (likely) more driving sources.

A number of mechanisms have been proposed to drive turbulence,
including \ion{H}{2} region expansion, supernova explosions, and fluid
instabilities involving magnetic fields and gravity (see
\citealt{mac04,elm04} for recent reviews).  Although less well
recognized, galactic spiral shocks are also an appealing means to
generate ISM turbulence.
\citet{woo76} showed that cloud deformation by passage of a shock, and
the associated Kelvin-Helmholtz instabilities, together trigger random motions
in the cloud.  \citet{wad04} suggested that an in-plane ``wiggle
instability'' of spiral shocks can also drive turbulent motions,
although this appears to be suppressed when three-dimensional
effects are included \citep[hereafter Paper I]{kim06}.  Very recently,
\citet{bon06} and \citet{dob06} demonstrated that passage of a distribution 
of clouds through a spiral shock gives rise to internal turbulent motions that
follow Larson's (1981) empirical scaling law fairly well.

Most of the studies cited above investigate two-dimensional dynamics driven 
by spiral shocks, neglecting the vertical degree of freedom.  When the
vertical dimension is taken into account and well-resolved, Paper I showed 
that non-steady motions of the spiral shock develop.\footnote{The early
low-resolution simulations of \citet{tub80} did not show this effect.}  
Non-steady behavior
of spiral shocks and generation of vertical motions has in fact been
seen in previous numerical models by other authors (e.g.,
\citealt{mar98,gom02,gom04,bol06}).  In this Letter, we clarify the
physical causes of non-stationarity for spiral shocks in stratified
disks.  We also quantify the level of the induced gas motions to show
that vertical spiral shocks should indeed be an important source of
turbulence in spiral galaxies.

\section{Methods and Model Parameters}

\begin{deluxetable*}{lccccccccc}
\tabletypesize{\footnotesize} 
\tablecaption{Summary of model parameters and simulation results 
\label{tbl-model}}
\tablewidth{0pt}
\tablehead{
\colhead{\begin{tabular}{c} Model                 \\ (1) \end{tabular}} &
\colhead{\begin{tabular}{c} $Q_0$                 \\ (2) \end{tabular}} &
\colhead{\begin{tabular}{c} $\beta_0$             \\ (3) \end{tabular}} &
\colhead{\begin{tabular}{c} $F$                   \\ (4) \end{tabular}} &
\colhead{\begin{tabular}{c} Grid                  \\ (5) \end{tabular}} &
\colhead{\begin{tabular}{c} $H$ (pc)              \\ (6) \end{tabular}} &
\colhead{\begin{tabular}{c} $\langle\sigma_x^2\rangle^{1/2}/\cs$  \\ (7) \end{tabular}} &
\colhead{\begin{tabular}{c} $\langle\sigma_y^2\rangle^{1/2}/\cs$  \\ (8) \end{tabular}} &
\colhead{\begin{tabular}{c} $\langle\sigma_z^2\rangle^{1/2}/\cs$  \\ (9) \end{tabular}} &
\colhead{\begin{tabular}{c} $\mathcal{M}_{\rm eff}$  \\ (10) \end{tabular}}
}
\startdata
A  & 1.8 & $\infty$ & 5  & $1024^2$ &196 & 0.66 & 0.63 & 0.38 & 4.0\\
B  & 1.8 & $\infty$ & 5  & $512^2$  &196 & 0.67 & 0.62 & 0.36 & 3.9\\
C  & 2.0 & $\infty$ & 7  & $512^2$  &218 & 1.01 & 0.86 & 0.45 & 5.0\\
D  & 2.5 & $\infty$ & 10 & $512^2$  &272 & 1.55 & 1.20 & 0.58 & 6.2\\
E  & 1.5 & 10       & 5  & $512^2$  &169 & 0.64 & 0.52 & 0.31 & 3.8\\
F  & 1.8 & 10       & 7  & $512^2$  &203 & 0.93 & 0.79 & 0.42 & 4.4\\
G  & 2.0 & 10       & 10 & $512^2$  &225 & 1.54 & 1.19 & 0.52 & 5.4
\enddata
\tablecomments{Col.\ (1)-(4): Model name and input parameters.
Col.\ (5): Numerical resolution.  
Col.\ (6): Vertical scale height of gas in
initial magnetohydrostatic equilibrium. 
Cols.\ (7)-(9): Density-weighted late-time velocity dispersions.
Col.\ (10): Effective Mach number of the large-scale spiral shock.}
\end{deluxetable*}

We study evolution of vertically-stratified gas flow across spiral
shocks in local regions of self-gravitating, differentially rotating,
magnetized galactic disks.  Our studies use a modified version of the
ZEUS code \citep{sto92a,sto92b} to solve the time-dependent
magnetohydrodynamic (MHD) equations presented in Paper I.  We use the
same local formulation described in Paper I, which explored fully
three-dimensional models.  Here, we perform much higher resolution
models to focus on the detailed turbulent response, but suppress the
degree of freedom parallel to the spiral arm (i.e. our models are
quasi-axisymmetric).  The reader is referred to Paper I for a complete
description of our model prescription and numerical methods.  Here, 
we briefly summarize the coordinate system and model
parameters we adopt.

We consider a local region centered on a tightly-wound spiral arm
(pitch angle $i\ll1$), with a potential perturbation due to the stars
assumed to be rigidly rotating with pattern speed $\Omega_p$.  We
introduce a local, co-rotating Cartesian frame centered on
$(R,\phi,z)=(R_0, \Omega_pt, 0)$.  The local frame is tilted such that
$\bf\hat x$ and $\bf\hat y$ denote the in-plane directions
perpendicular and parallel, to the arm, respectively, while $\bf\hat
z$ is perpendicular to the galactic plane \citep{rob69}.  We set up a
two-dimensional simulation domain with size $L_x \times L_z$ in the
$x$-$z$ plane (hereafter XZ plane), and assume all physical variables
to be independent of $y$.  The simulation box size $L_x$ is equal
to the arm-to-arm separation. We allow for nonzero values of velocity
$v_y$ and magnetic field $B_y$; 
initially $v_y = (\Omega_0 - \Omega_p)R_0 - \Omega_0 x$ and 
$B_y$ is independent of $x$, where $\Omega_0$ is the local angular velocity
at $R_0$. We also include Coriolis forces. 
This quasi-axisymmetric approximation
prevents clump-forming, non-axisymmetric gravitational instabilities
from developing, enabling us to focus on the properties of turbulence
independent of cloud formation.

We idealize the ISM by treating it as an isothermal gas with effective
speed of sound $\cs=7\kms$.  The initial gaseous disk (without the
spiral potential) is in vertical magnetohydrostatic equilibrium.  As
reference values, we take $R_0=10\kpc$, $\Omega_0=26\kms\kpc^{-1}$, 
and epicyclic frequency $\kappa_0=
2^{1/2}\Omega_0$ (for a flat rotation curve).  The corresponding
orbital period is $\torb \equiv 2\pi/\Omega_0 = 2.4 \times 10^8 {\rm
 yr}$.  For the spiral arm parameters, we adopt $\Omega_p
=\Omega_0/2$ and $\sin i=0.1$.  We adopt $L_x=\pi R_0 \sin i
=3.14\kpc$ for a two-armed spiral.  We impose reflection symmetry with
respect to the midplane, and apply open boundary conditions at
$z=L_z=4H$, where $H$ $(\sim170-270$ pc) is the initial disk scale height.
We adopt sheared-periodic boundary conditions in $x$.

The three key parameters that characterize our model disks are
\begin{equation}
Q_0= \frac{\kappa_0 c_s}{\pi G \Sigma_0}, 
\;\;
\beta_0 = \frac{c_s^2}{\vA^2},
\;\;
F=\frac{2}{\sin i}\left(\frac{|\Phi_{\rm sp}|}{R_0^2\Omega_0^2}\right),
\end{equation}
where $\Sigma_0$ is the gas surface density, $\vA$ is the Alfv\'en
speed, and $\Phi_{\rm sp}$ is the amplitude of the imposed sinusoidal 
potential (Paper I).  The quantity
$F$ measures the ratio of the perturbed sinusoidal radial force to the mean
axisymmetric gravitational force \citep{rob69}.  We do not 
consider the vertical variation of the spiral potential perturbation; 
the perturbed vertical force is negligible ($<4\%$) compared to the 
background vertical force.
In our simulations,
the potential amplitude slowly increases from $0$ to $F$ over
$\sim1.5\torb$.  We present results for seven numerical models, as listed in
Table~\ref{tbl-model}.

\section{Nonlinear Simulations}

\subsection{Spiral Shock Evolution in a Stratified Disk}

\begin{figure}
\epsscale{1.0}
\plotone{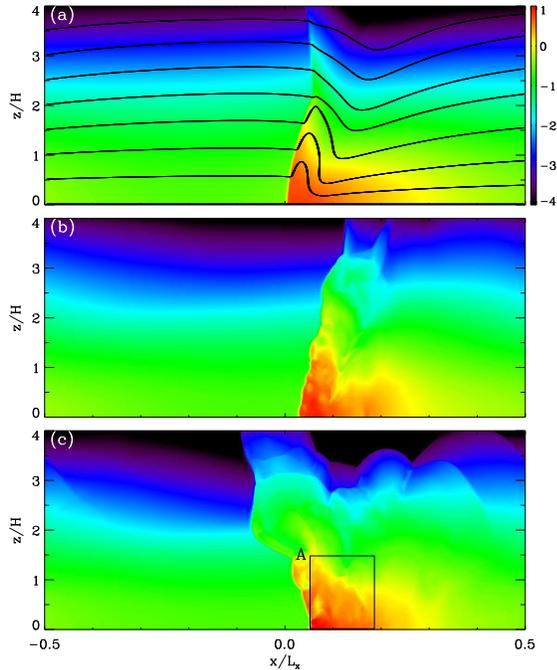}
\caption{Snapshots of density in logarithmic color scale of model G at
$t/t_{\rm orb} = 1.2, 1.7, 1.8$ from top to bottom.
Solid lines in (a) represent instantaneous streamlines of gas.
The rectangular box A in (c) indicates a sector enlarged 
in Figure~\ref{fig-vel}.
\label{fig-snap}}
\end{figure}

In this subsection, we describe the evolution of spiral shocks in our fiducial 
model G;  evolution in other models is similar.
As the amplitude of the stellar spiral potential grows, magnetosonic waves
emerge and steepen, forming a shock front near $x/L_x=0.04$ 
by $t/\torb\sim 1.0$.
At this time, fluid motions are nearly horizontal, and the shock front is
nearly vertical.  The gravity due to the large post-shock midplane density 
enhancement soon breaks vertical force
balance, however, pulling material at high $|z|$ toward the midplane.  
With increased postshock density, the shock near $z=0$ tends to move
upstream toward the spiral potential minimum, causing the shock front 
to bend.  Gas entering the curved shock is initially bent upward
toward the shock front and then pulled downward due to
strong vertical gravity.  The falling material establishes 
repulsive pressure gradients and bounces back to high $|z|$.  
The instantaneous streamlines shown for model G at $t/\torb= 1.2$ in
Figure~\ref{fig-snap}{\it a} reflect gas motions at this stage.

Our model simulations show that spiral shocks in the XZ plane never
achieve a steady state, as illustrated by the strong density
variations in Figure~\ref{fig-snap}{\it b,c} for model G.  This
result is quite unlike the steady solutions that obtain for models
which neglect vertical degrees of freedom (e.g.,
\citealt{woo75,kim02}).  In particular, 
when the fluid variables are allowed to vary
with $z$, the gas leaving the right $x$-boundary does not in general
have the same $z$ as it had when it originally entered the left
$x$-boundary\footnote{This is because the fundamental vertical
  oscillation period need not equal the arm-to-arm flow period,
  although these timescales are typically within a factor of a few of
  each other.}, as Figure~\ref{fig-snap}{\it a} illustrates. When the
vertical gravity exceeds the vertical repulsive pressure force
downstream from the shock, post-shock material is drawn toward the
midplane, and the shock front at high $|z|$ bends downstream (Fig.\
\ref{fig-snap}{\it b}).  As gas is further compressed, vertical
pressure gradients overwhelm vertical gravity and the gas even at
$|z|/H \sim 0.5$ is able to rebound to $|z|/H \simgt 3$.  This in turn
forces the shock front at large $|z|$ to shift back upstream (Fig.\
\ref{fig-snap}{\it c}).  The vertically-rebounding gas overshoots
equilibrium, vertical gravity again dominates, and the cycle repeats.
As gas traverses the spiral arms, therefore, the shocks in our models
continue to quasi-periodically shift back and forth perpendicular to
the arms.  These ``flapping'' motions of spiral shocks are strongest
at high $|z|$. Random motions at scales comparable to the height $H$
are also driven, and many weak shocks develop especially at high $|z|$
(see Fig.\ \ref{fig-snap}{\it c}).

In addition to large-scale flapping, we find that spiral shocks in
model G experience two small-scale instabilities that aid in transfer
of random gaseous kinetic energy to smaller scales: an
advective-acoustic cycle and vortex generation.  The
advective-acoustic cycle occurs as the most overdense midplane region
perturbs the upstream flows by launching acoustic waves, analogous to
the vortical-acoustic cycle for instabilities in isothermal
Bondi-Hoyle-Lyttleton accretion (e.g., \citealt{fog02}).  Because the
spiral shock front is curved in the XZ plane, Crocco's theorem ensures
generation of vorticity.  Moreover, strong rising and falling currents
of the gas in the post-shock region create a sheared velocity field
favorable to Kelvin-Helmholtz instability, which can also generate
vorticity.  We remark that vortices formed at the shock front are not
fully resolved in our models because instabilities grow faster at smaller
scales.  Nevertheless, the velocity
dispersions are essentially independent of numerical resolution (see
below), suggesting that these instabilities may enhance the small-scale 
cascade but do not affect the overall level of induced
turbulence.

\begin{figure}
\epsscale{1.00}
\plotone{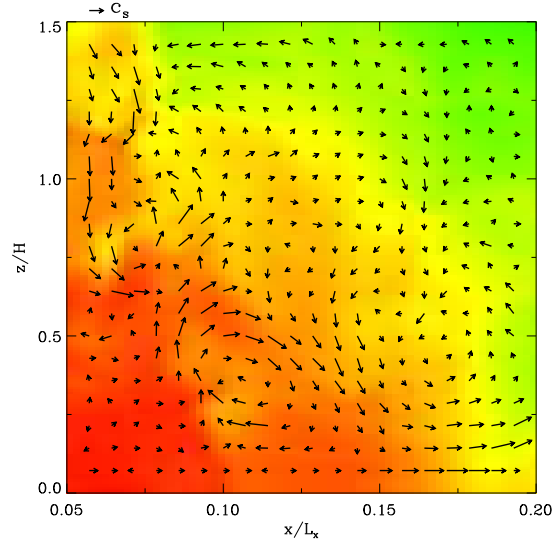}
\caption{Velocity vectors and density in logarithmic scale at 
$t/\torb=3.2$ in the region A marked in Figure \ref{fig-snap}c.
The color scale is the same as in Figure \ref{fig-snap}. The size of
the arrow above the box corresponds to the sound speed.
\label{fig-vel}}
\end{figure}

Figure \ref{fig-vel} plots typical velocity and density structures inside
a spiral arm when turbulence is saturated.  
Many weak shocks and well-resolved vortices of both signs are apparent.
The instantaneous, density-weighted, total velocity dispersion in the
region shown in Figure \ref{fig-vel} is 1.51 times the sound speed.
We find that the power spectra of the perturbed velocities
extend smoothly over all scales, indicating fully-developed
turbulence.  Because of the large-scale shock in $v_x$, the large 
shear/streaming in $v_y$, and
the vertical stratification, the power spectrum of $v_z$ as a function
of $k_x$ best characterizes the true turbulence. 
At heights $|z|/H<1$, this ranges from 
$v_z^2(k_x) \propto k_x^{-0.5}$ at $n_x\equiv L_x k_x/(2\pi)< 8$ to 
$v_z^2(k_x) \propto k_x^{-0.8}$ at $8 < n_x <16$ to
$v_z^2(k_x) \propto k_x^{-1.3}$ at $16 < n_x <64$ for
the $k_z=0$ modes. 
For $1<|z|/H<2$, the power spectra are slightly steeper, having
$v_z^2(k_x) \propto k_x^{-0.6}$ at $n_x < 8$, 
$v_z^2(k_x) \propto k_x^{-1.1}$ at $8 <n_x <16$, and
$v_z^2(k_x) \propto k_x^{-1.7}$ at $16 <n_x <64$.

\subsection{Level of ISM Turbulence}

\begin{figure}
\epsscale{1.0}
\plotone{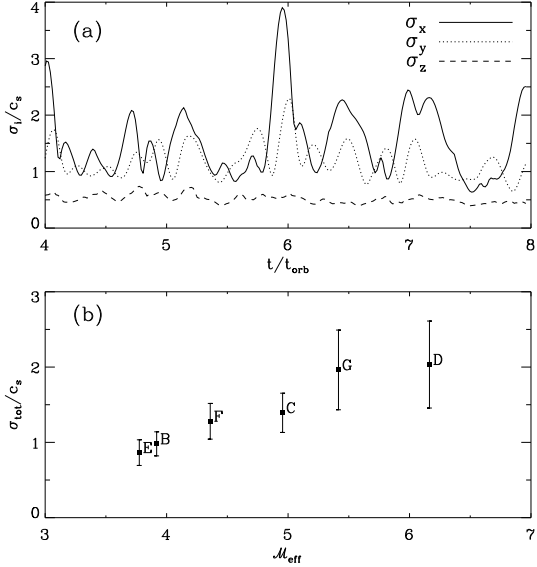}
\caption{({\it a}) Time evolution of the density-weighted velocity 
dispersions $\sigma_i$ in model G.
({\it b}) Total velocity dispersion $\sigma_{\rm tot}$ {\it vs.}
the effective Mach number $\mathcal{M}_{\rm eff}$ of the spiral shock
in models B to G.
The errorbars represent the standard deviations
in the temporal fluctuations of $\sigma_i$.
\label{fig-sig}}
\end{figure}

We now quantify the level of random gas motions driven by spiral
shocks.  Since the velocities in XZ spiral shocks are non-uniform and
non-stationary, it is useful to construct the mean velocity field
$\langle v_i\rangle$ (with $i=x$, $y$, or $z$), where
the bracket $\langle\;\rangle$ denotes a time average over
$t/\torb\sim 4-8$ after turbulence saturates.  We then measure 
density-weighted velocity dispersions using
$\sigma_i^2 \equiv {\int \rho\delta v_i^2 dx dz}
/{\int \rho dx dz},$
where $\delta v_i\equiv v_i - \langle v_i\rangle$.\footnote{In
computing each $\sigma_i$, if we initially subtract out 
velocities corresponding either to unperturbed rotation or to a 
thin-disk spiral shock (instead of $\langle v_i\rangle$), then the
velocity dispersions would be larger.}  
Figure~\ref{fig-sig}({\it a}) plots $\sigma_i(t)$ for model G, 
while columns (7)-(9)
in Table~\ref{tbl-model} list $\langle \sigma_i^2 \rangle^{1/2}$ for
all the models.

As Figure~\ref{fig-sig}({\it a}) shows, random gas motions in model G are
supersonic in the $x$- and $y$-directions, and exhibit large-amplitude
temporal fluctuations.  The characteristic periods of these
quasi-periodic fluctuations are in the range $\sim0.5-0.9\torb$.
The maximum velocity dispersions occur when
the spiral shock is temporarily vertical.
Evidently, the system reaches a quasi-steady state in which
dissipation of turbulence (in shocks and through cascades) is offset
by the continual input of new turbulent energy from the large-scale
flapping.  Turbulence is strongest in the immediate post-shock region,
and declines with increasing distance from the shock.  
Vertically-averaged velocity dispersions
inside the spiral arms typically exceed those in interarm regions by about a
factor of 2.  For the model parameters we have considered,
Table~\ref{tbl-model} indicates that $\langle\sigma_x\rangle \sim
\langle\sigma_y\rangle \sim 2\langle\sigma_z\rangle$.  Comparison of
$\langle\sigma_i\rangle$ between models A and B shows that the
velocity dispersion is independent of numerical resolution.

What determines the level of velocity dispersions in XZ spiral shocks?  
Table~\ref{tbl-model} suggests that stronger shocks (having larger
values of $Q_0^{-1}$, $\beta_0$, and/or $F$) yield larger $\sigma_i$. 
To characterize the shock strength in each model as simply as possible, we 
have run one-dimensional counterparts with the same set of
parameters, using the thick-disk self-gravity prescription (see Paper I).
We define the effective Mach number 
$\mathcal{M}_{\rm eff}\equiv 
\langle\Sigma_{2} / \Sigma_{1}\rangle^{1/2}$,
where $\Sigma_1$ and $\Sigma_2$ denote
the preshock and postshock surface densities, respectively, 
in the resulting (one-dimensional) spiral shock.
Column (10) in Table~\ref{tbl-model} lists $\mathcal{M}_{\rm eff}$ for
each model. 
Figure~\ref{fig-sig}({\it b}) plots the total velocity dispersion 
$\sigma_{\rm tot} = 
(\langle\sigma_x^2\rangle + \langle\sigma_y^2\rangle + 
 \langle\sigma_z^2\rangle)^{1/2}$ as a function of $\mathcal{M}_{\rm eff}$. 
While $\sigma_{\rm tot}$ monotonically increases roughly
as $\sigma_{\rm tot}/\cs = 0.6 \mathcal{M}_{\rm eff} - 1.4$
for $\mathcal{M}_{\rm eff} < 5.5$, 
it is more or less constant at about $2\cs$ for $\mathcal{M}_{\rm eff}>5.5$.
We have found that models with other values of $\Omega_p/\Omega_0$ follow 
the same $\sigma_{\rm tot}$--$\mathcal{M}_{\rm eff}$ relation as shown in 
Figure~\ref{fig-sig}({\it b}). 

\section{Discussion}

Galactic rotation can supply an effectively inexhaustible amount of
kinetic energy to power turbulence in the ISM
\citep{fle81}.\footnote{To maintain a fixed level of turbulence, 
the energy dissipation per unit mass 
per unit time must be balanced by the net stresses, 
$\Omega \langle\rho v_x \delta v_y - B_x B_y/(4\pi)\rangle/\bar\rho$ 
(e.g. \citealt{pio05}).  
If this power is supplied at the expense of overall ISM accretion
toward the galactic center, the accretion time is 
$\sim (2\Omega)^{-1} \bar\rho (\Omega R)^2 \langle \rho v_x
\delta v_y \rangle^{-1}$,  which exceeds
the Hubble time if $\langle \rho v_x \delta v_y - B_x B_y/(4\pi)
\rangle \sim \bar \rho c_s^2$.}
In this Letter, we have shown that spiral shocks are in general
non-stationary in the radial-vertical plane, and can efficiently
transform some of the available bulk rotational energy into random gas
motions.  Vertical force imbalance drives radial flapping of the shock
front at $|z|/H \simgt 0.2$, which then feeds turbulence at smaller
scales.  The random gas motions induced by the spiral shock persist
despite strong shock dissipation, and yield time-averaged in-plane
velocity dispersions $\sim 7-10\kms$ for a range of shock strengths,
similar to the observed line widths of cold and warm atomic gas in the
Milky Way (e.g., \citealt{hei03}).  Vertical velocity dispersions are
lower, but still amount to $\sim 1/2$ of the thermal velocity
dispersion.

Although our presentation has focused on shocks with realistic
parameters, in fact turbulence generation by spiral shocks appears to
require neither self-gravity nor magnetic fields.
Our model simulations, including those unlisted in Table~\ref{tbl-model},
show that anything (e.g., an imposed variation of
$\Phi_{\rm sp}$ with $z$) that makes the primary shock non-vertical will
end up producing a non-steady flow -- simply because vertical
oscillation periods need not (and in general do not) agree with the
horizontal crossing time between arms. 

Since strong spiral arms usually imply a high rate of star formation,
spiral shocks and stellar energy sources may work together in
generating turbulence in many galaxies.  The absence of observed correlation
between spiral arm phase and turbulent amplitude (e.g.,
\citealt{dic90}), however,  suggests that these two processes are not
the only important sources of turbulence.  In addition, radio
observations of extended \ion{H}{1} disks in face-on
galaxies show that the total vertical velocity dispersions are as
large in the non-star-forming outer parts as in the star-forming inner regions 
\citep{van99}, again suggesting that additional sources of turbulence
are able to compensate when needed.

One compensating source of turbulence may be provided by the
magnetorotational instability (MRI) \citep{sel99,kim03,dzi04}.  The MRI
should naturally be strongest exactly where spiral shocks are absent:
in the outer galaxy, where gravity is weak, the disk flares, and the
density drops; and in interarm regions, where the angular velocity
decreases outward (in arms, shear is reversed so that MRI cannot
occur).  

While in this Letter we adopt isothermal conditions for the gas and
take the mean magnetic field parallel to the spiral arm, the real ISM
has a multi-phase structure and is threaded also by weak vertical
magnetic fields.  \citet{pio04,pio05} demonstrated that MRI is able to
generate random gas motions up to $\sim 8\kms$ in two-phase gas under
favorable conditions.  It will be interesting to see how turbulent
driving in spiral shocks develops for a multiphase medium, and whether
MRI and spiral shocks can indeed establish a
geographically-compensating balance of power.

\acknowledgments

We are grateful to S.\ S.\ Hong, J.\ Kim, and D.\ Ryu for valuable 
discussions, and to the referee I. Bonnell for a helpful report.  
This work was supported partly by Korea Science and Engineering
Foundation (KOSEF) grant R01-2004-000-10490-0 and partly by
NASA under grant NNG05GG43G.

\end{document}